\documentclass[11pt,twoside]{article}

\usepackage{graphicx,amssymb}

\begin{document}
% ---------------------------------------------------------------------
%
%-------------------------------------------------------------
%
%
\thispagestyle{empty}

\leftline{\copyright~ 1998 International Press}
\leftline{Adv. Theor. Math. Phys. {\bf 2} (1998) 987-1009}  

\vspace{0.4in}
\begin{center}
{\huge \bf Duality without supersymmetry}

\vspace{0.4in}

\({\bf  Paul\;   Fendley}\)
\linebreak
Physics Department\\
University of Virginia\\
Charlottesville, VA 22901\\

%\vspace{0.2in}

%\title{Duality without supersymmetry}

%\author{Paul Fendley}
%\maketitle
%\centerline{Physics Department}
%\centerline{University of Virginia}
%\centerline{Charlottesville, VA 22901}
\vspace{0.2in}
{\bf Abstract} \\
\vspace{0.1in}
%\begin{abstract}

\parbox[c]{4.5in}{\small \hspace{.2in}I show that physical quantities in several two-dimensional
condensed-matter models are related to the Seiberg-Witten calculation
of exact quantities in supersymmetric gauge theory. In particular, the
magnetization in the Kondo problem and the current in the boundary
sine-Gordon model can each be expressed in the form $\int dx/y$, where
for example in the latter $y^2 = x + x^g - u^2$ with $u$ related to
the boundary mass scale (the analog of $\Lambda_{\rm QCD}$) and
$g\propto R^2$, where $R$ is the radius of the boson. Thus for
irrational $g$, the hyperelliptic curve $y(x)$ is of infinite genus,
while for rational $g$ it is of finite genus. The models are
integrable and possess a quantum-group symmetry for any $g$, but are
supersymmetric only at $g=2/3$. Both models also possess unique forms
of $g\to 1/g$ duality.}
\end{center}

\renewcommand{\thefootnote}{}
\footnotetext{\small e-print archive: {\texttt http://xxx.lanl.gov/abs/hep-th/9804108}}
\renewcommand{\thefootnote}{\arabic{footnote}}

\section{Introduction}

Several years ago, Seiberg and Witten showed that ``duality'' can
imply a much more elaborate structure than merely invariance under
inverted couplings. In $N$=2 supersymmetric gauge theories, they
showed that one can exploit the analyticity implied by the
supersymmetry in order to do non-perturbative computations
\cite{SW}. This work has been generalized in many ways, but always to
field theories with supersymmetry, since analyticity in some parameter
is required. However, there seems to be no fundamental reason why
supersymmetry is required for such analyticity to be present. In this
paper, I will show that several fundamental models of integrable \newpage
\noindent$1+1$-dimensional field theory which are not supersymmetric exhibit
similar behavior.
\pagenumbering{arabic}
\setcounter{page}{988}

\pagestyle{myheadings}
\markboth{\it DUALITY WITHOUT SUPERSYMMETRY}{\it P. FENDLEY }
A familiar fact from complex analysis is that if one knows the poles,
residues and asymptotic behavior of a function, then that function can
be reconstructed uniquely if it is analytic except at the
poles. Similarly, if a function has branch-point singularities but is
analytic elsewhere, one can reconstruct the complete function by
knowing enough about the behavior at the singularities. In the
simplest non-trivial cases, the function $f(u)$ can be expressed in
terms of integrals like
\begin{equation}
f(u) = \int_{\cal C} \frac{dx}{y}
\label{ellip}
\end{equation}
for some contour ${\cal C}$. Here $y^2$ is a cubic polynomial in
$x$ with coefficients depending on $u$; by redefining $x$ one can
always write it in the form
$$
y^2= x^3 + a(u)x + b(u)\ .
$$
The integrand has three square-root branch points at the roots of this
polynomial, so there are two different branch cuts in the $x$ plane,
one connecting two of the branch points, and the other connecting the
third to infinity. The sign ambiguity in $y$ can be fixed by allowing
$x$ to take values on two sheets. Including the point at infinity,
each sheet is a sphere with two branch cuts. The two spheres must be
glued together along both branch cuts; this is 
equivalent to a torus. For higher-order polynomials in $x$, there are
more singularities and the corresponding surface has higher genus.
At special values of $a$ and $b$ (i.e.\ special values of $u$) where
two of these roots coincide, the integral (\ref{ellip})
logarithmically diverges if the contour goes in between these two
roots. These values of $u$ are the singularities of $f(u)$, and are
values of $u$ where one of the cycles of the surface is pinched to a
point.

In \cite{SW}, Seiberg and Witten showed that certain quantities in
$N$=2 supersymmetric $SU(2)$ gauge theory are holomorphic functions of
$u$, an order parameter related to the vacuum expectation value of the
Higgs field. The ``duality'' is the fact that the monodromies around
the singularities in the $u$ plane are given by $SL(2,Z)$
transformations generalizing $g\to 1/g$. These monodromies can exactly
be determined by perturbation theory and an exact mass formula for
certain kinds of particles called BPS states. Thus various physical
quantities (for example the effective coupling constant, the beta
function and the mass gap) are given {\it exactly} and {\it
non-perturbatively} by integrals like (\ref{ellip}). Their work has
now been generalized to many supersymmetric gauge theories, and also
is deeply related to duality in string theory. There are many review
articles on these subjects; see for example \cite{SWreview}.

In integrable models, it has long been known how to exploit
analyticity to derive various physical quantities from information
like functional relations. In this paper I will show that physical
quantities in several integrable models of $1$+$1$-dimensional
boundary quantum field theory can be expressed as integrals like
(\ref{ellip}). The theories, the Kondo model and the boundary
sine-Gordon model, are not supersymmetric, and it is not likely that
there is a hidden supersymmetry. Here the analyticity seems to follow
from the integrability or, equivalently, the quantum-group symmetry of
the models. The distinction between the models discussed here and
integrable models where analyticity has previously been utilized is
that these models obey intriguing, and I believe unique, forms of
$g\to 1/g$ duality \cite{FW,FLSbig}. This sort of duality, which
generalizes the electric-magnetic duality of electromagnetism, was
observed long ago in some gauge theories \cite{mono}, providing one of
the motivations for the work of Seiberg and Witten. This paper thus
explores two somewhat different forms of duality: the ability to
express physical quantities as curves around branch-point
singularities, and the ability to relate different physics by a $g\to
1/g$ mapping.

While the main purpose of this paper is to discuss the Kondo and
boundary-sine Gordon models as surprising manifestations of
Seiberg-Witten theory, it is worthwhile noting that both are extremely
interesting in their own right. Not only are they fundamental models
of strongly-interacting statistical mechanics, but both have been
observed experimentally. In particular, the Kondo model used
successfully for decades to describe dilute magnetic impurities in a
conductor (see e.g. \cite{TW,AFL}) while the boundary sine-Gordon
model describes the current-carrying edge excitations tunneling
through a point contact in a fractional quantum Hall device (see
\cite{Hall}). The duality in the latter maps Laughlin quasiparticles
to electrons.

In section 2 I introduce the Kondo model and recall various important
results from the Bethe ansatz computations. Section 3 contains the
main result of the paper, the writing of the magnetization in the
Kondo problem as an integral over a cycle of a curve. This form is
equivalent to the Bethe ansatz result, but much simpler and more
useful. For example, it gives the entire singularity structure of the
magnetization. Section 4 discusses a peculiar duality present in the
Kondo and boundary sine-Gordon models. Section 5 highlights the
$SU(2)$ invariant limit of the Kondo problem, which physically is the
most reminiscent of QCD.

\section{The Ingredients}

In this section I introduce the Kondo model, and discuss the
ingredients necessary for understanding its type of Seiberg-Witten
duality.

The Kondo model is one of the fundamental models of statistical
mechanics. It describes three-dimensional non-relativistic electrons
coupled to a single impurity spin. Writing the electron wavefunction
in terms of spherical harmonics around the impurity, the most relevant
interaction comes from the $s$-waves. Since these are spherically
symmetric, the results depend only on the radial coordinate, and the
problem is effectively $1+1$-dimensional: space is a half-line, with
the impurity located at the boundary. Moreover, in the one-dimensional
effective model, the fermions are relativistic and massless because
one is concerned only with excitations near the Fermi surface. When
$p$ is the momentum difference from the Fermi momentum $p_F$ and $E$
the energy difference from the Fermi energy $p_F^2/2m$, $E+E_F=
(p+p_F)^2/2m \approx p_F^2/2m + v_F p$. Therefore the relativistic
dispersion relation $E=pv_F$ holds, with the Fermi velocity
$v_F=p_F/m$ playing the role of the speed of light. Henceforth
$v_F=1$. Since there are two spins of electrons in the original
three-dimensional problem, there are two flavors of fermions in the
effective one-dimensional problem. One can then form an $SU(2)$
``spin'' current ${\bf J}= \psi^{\dagger i} {\bf \sigma}_{ij} \psi_j$
using the Pauli matrices ${\bf \sigma}$, and the $U(1)$ ``charge''
current ${\cal J}=\psi^{\dagger i}\psi_i$. In conformal field theory
language, these yield $SU(2)_1$ and $U(1)$ WZW theories respectively,
each with a central charge $c=1$ \cite{AL}.

The impurity is represented by a quantum-mechanical spin ${\bf S}$ in
the spin-$S$ representation. An impurity located at $x=0$ is coupled
antiferromagnetically to the fermions by a term in
the Lagrangian $-\lambda {\bf J}(0)\cdot {\bf S}$ for positive
$\lambda$. There is no integral in this term because the interaction
takes place at a single point in space, the boundary of the half-line.
The coupling $\lambda$ is dimensionless since
the current is of dimension one, so naively it seems that this
coupling should be marginal and preserve conformal symmetry at the
boundary.  However, Kondo observed more than three decades ago that if
one does perturbation theory in $\lambda$, there is a short-distance
divergence \cite{Kondo}.  Thus the interaction term is relevant, and a
mass scale is present in the theory. In particle-physics language, the
Kondo model is asymptotically free and undergoes dimensional
transmutation.  This scale generated is usually called the Kondo
temperature $T_K$, and it is completely analogous to $\Lambda_{QCD}$
in gauge theory.  In terms of the original parameter $\lambda$,
\begin{equation}
T_K \sim \lambda^{1/2}e^{-const/\lambda}.
\label{tklam}
\end{equation}
The precise relation as well as the positive constant are not
universal (i.e.\ depend on regularization procedure); we will discuss
physics in terms of $T_K$ and will not need the detailed definition.

As $\lambda$ gets large (or more precisely, we study physics at energy
scales below $T_K$), the system crosses over to a strongly-coupled
phase.  At $T_K\to\infty$, there is another fixed point, where the
spin is screened by one of the electrons (because of Pauli exclusion
only a single electron can bind to the impurity). At this
strongly-coupled fixed point, the spin of the impurity is effectively
reduced from $S$ to $S- 1/2$. The physical quantity we will study is
the magnetization ${\cal M}_S$ of the spin-$S$ impurity as a function of
applied magnetic field $H$. At zero temperature, the ${\cal M}_S$ is a
function only of the dimensionless quantity $H/T_K$, so
${\cal M}_S(\infty)=S$, while $M(0)=S - 1/2$.

Since Kondo's work, the Kondo model has been analyzed using many
different techniques. It was one of the first models to be thoroughly
understood using scaling and renormalization-group techniques.
Various properties of the strongly-coupled fixed point were understood
using scaling arguments \cite{And}. The crossover between the two
fixed points was described using numerical renormalization-group
techniques \cite{Wilson}. Subsequently, many quantities were computed
exactly using the Bethe ansatz; for reviews see \cite{TW,AFL}. The
exact result for the magnetization curves at zero temperature can be
written in the form \cite{FW}
\begin{eqnarray}
{\cal M}_S\left(\frac{H}{T_K}\right)&=& S- \frac{1}{2} + \frac{i}{8\sqrt{\pi}}
\int_{-\infty}^\infty
d\omega\ e^{-2i\omega\ln(H/T_K)}
\frac{\Gamma(1/2 + i\omega)}{\omega+i\epsilon}\nonumber \\
&& \times[f_+(\omega)]^{2S} [f_-(\omega)]^{2S-1}
\label{magiso}
\end{eqnarray}
where
$$f_\mp(\omega) \equiv
\left[\frac{\pm i\omega+\epsilon}{2\pi} \right]^{\pm i\omega}$$
with $\epsilon$ positive and tending to zero. The purpose of this
paper is to write an equivalent expression for ${\cal M}_S$ which not only
simplifies matters but which also yields a connection to the gauge-theory
results of Seiberg and Witten.

I will discuss a more general model, the anisotropic Kondo model,
which allows for $SU(2)$-breaking interactions with the impurity, namely
$$\lambda {\bf J\cdot S} +\lambda_z J_z S_z.$$ The model with
anisotropy arises in dissipative quantum mechanics, where it describes
a particle moving in a double well with dissipation \cite{Legg}. A
convenient way of parameterizing the anisotropy is given by bosonizing
the model; in fact the Kondo model was one of the first models to be
bosonized \cite{Schotte}.  The $U(1)$ charge current ${\cal J}$ does
not couple to the impurity, so it can be ignored.  The bulk theory is
described by a single free boson, with Lagrangian taking the form
\begin{equation}
L_0=\frac{1}{4\pi g}\int_0^\infty dx
(\partial_\mu\phi)^2 -
\frac{H}{2\pi g}\int_0^\infty
dx\ \partial_t\phi.
\label{lzero}
\end{equation}
In the usual conventions of conformal field
theory, the boson takes values on a circle so that
$\phi=\phi+2\pi R$. Changing the coupling $g$ is equivalent to
rescaling the boson and changing the radius of this circle;
i.e. $g\propto R^2$. In this paper it will be convenient to stick with
the coupling $g$. In the bosonized language, the coupling to the
impurity is then
\begin{equation}
L_S=\lambda_g\left(S_+e^{-i\phi(0)}+
S_-e^{+i\phi(0)}\right)-H S_z.
\label{lkondo}
\end{equation}
The Lagrangian $L_0+ L_S$ describes the Kondo model when $g\le 1$.
The anisotropy is parametrized by $g$, which is defined so that the
isotropic case has $g=1$.  One well-studied value (known as the
Toulouse limit) is $g=1/2$, where the problem can be mapped on to one
of free fermions. The magnetic field $H$ couples to the conserved
$z$-component $S_z + \int J_z$ of the spin.  The operators $\exp(\pm
i\phi(0))$ have boundary scaling dimension $1$ when $g=1$ as they
should; in general they have dimension $g$. Therefore
$$T_K\propto \lambda_g^{1/(1-g)}$$
for $g\ne 1$. For a microscopic
definition of $T_K$ (i.e.\ including the cutoff), see \cite{TW}.

Traditionally in the Kondo problem, one takes the matrices $S_i$ to
act in the spin-$S$ representation of $SU(2)$. However, a subtlety
arises for $g< 1 $ when spin is greater than $1/2$.  For the problem
to be integrable, one must instead take the matrices to act in the
spin-$S$ representation of the quantum-group algebra $SU(2)_q$ instead
of the ordinary $SU(2)$ algebra \cite{FLeS}. The quantum group is
actually an algebra defined by the relations
$$[S_z,S_\pm]=\pm 2S_\pm,\quad [S_+,S_-]=\frac{q^{S_z}-q^{-S_z}}{
q-q^{-1}},$$
where $q=e^{i\pi g}$ .  In the isotropic case $q=-1$ or in the
classical limit $q=1$, this reduces to the usual $SU(2)$ algebra.
The Pauli matrices satisfy this algebra for any $q$, so the
distinction between $SU(2)$ and $SU(2)_q$ is not important for
$S=1/2$. In the following, the anisotropic Kondo model of spin-$S$ is
{\bf defined} to be the model with the $q$-deformed algebra; it can be
identified with the ``physical'' Kondo model for $S=1/2$ or for $g=1$.

The magnetization for spin-$1/2$ has been found exactly using the
Bethe ansatz method. The dimensionless parameter $u$ is defined by
\begin{equation}
u\equiv
\frac{\Gamma(\frac{1}{2} + \frac{1}{2(1-g)})}{2\sqrt{\pi}
\Gamma(1+ \frac{1}{2(1-g)})}
 \frac{H}{T_K}.
\label{udef}
\end{equation}
The reason for the extra numerical factors in front is to conform with
the usual definition of $T_K$ \cite{TW}. The model is ``physical''
when $H/T_K$ or equivalently $u$ is real and positive, but for reasons
which will soon become apparent, we will study the magnetization for
all values of $u$ in the complex plane. The Bethe ansatz result for
the magnetization is \cite{Waniso}
\begin{equation}
M_{1/2}(u)= \frac{i}{4\pi^{3/2}}
\int_{-\infty}^\infty
d\omega\ e^{-2i\omega\ln(u)}
\frac{\Gamma(1/2 + i\omega)\Gamma(1-\frac{i\omega}{1-g})}
{(\omega+i\epsilon)\Gamma(1-\frac{ig\omega}{1-g})}.
\label{maganiso}
\end{equation}
In the limit $g\to 1$, $M_{1/2}$ here indeed reduces to the expression
${\cal M}_{1/2}$ in (\ref{magiso}).

One can also derive a similar integral expression for $M_S(u)$ for any
$g$, finding for example that although $M_S(\infty)=S$ as with the
isotropic case, $M_S(0)=(S-1/2)/g$.  Instead of using the integral
expressions for higher $S$, it will be much more useful to utilize
{\it fusion relations}. Fusion relations are a way of obtaining a new
integrable model from an existing one. For example, from the
spin-$1/2$ Heisenberg spin chain, one can obtain the integrable
spin-$1$ chain by fusing together two neighboring spins \cite{KRS}. In
the Kondo model, this procedure allows one to construct the
higher-spin impurity models by continuing the spin-$1/2$ model to
imaginary couplings. The partition functions
$Z_S(H/T_K,H/T)$ at arbitrary temperature obey\cite{BLZ}
\begin{equation}
Z_S\left(i\frac{H}{T_K},\frac{H}{T}\right)
Z_S\left(-i\frac{H}{T_K},\frac{H}{T}\right)
= 1 +Z_{S+1/2}\left(\frac{H}{T_K},\frac{H}{T}\right)
Z_{S-1/2}\left(\frac{H}{T_K},\frac{H}{T}\right)
\label{fusionBLZ}
\end{equation}
with $Z_0\equiv 1$.  These relations hold for any $g$. However,
the quantum-group algebra has the peculiar property that when $g=P/Q$
for mutually prime integers $P$ and $Q$, then $(S_\pm)^Q=0$.  Thus as
opposed to ordinary $SU(2)$, the representation of spin $Q/2$ is
reducible.  This means that the partition functions ``truncate'' when
$g$ is rational (i.e.\ $q$ is a root of unity) \cite{FLeS}:
\begin{equation}
Z_{Q/2} \left(\frac{H}{T_K},\frac{H}{T}\right) =
Z_{Q/2-1} \left(\frac{H}{T_K},\frac{H}{T}\right)  +
2 \cosh\left[Q\frac{H}{2T}\right]\qquad\qquad
g=\frac{P}{Q}\
\label{truncT}
\end{equation}
In the subsequent analysis, this truncation will explain why the
curves used to describe the magnetization have finite genus for $g<1$
rational, but infinite for $g=1$ or irrational.

This paper will be concerned exclusively with the physics at zero
temperature, so that the magnetization depends only on $H/T_K$ and
$g$. In this limit the fusion relations (\ref{fusionBLZ}) become
linear because the free energy $F_S=-T\ln Z_S$ remains finite as $T\to
0$. In terms of the magnetization $M_S=-\partial F_S/\partial H$, the
limit of (\ref{fusionBLZ}) is
\begin{equation}
M_S(iu) + M_S(-iu)= M_{S-1/2}(u) + M_{S+1/2}(u)
\label{fusion}
\end{equation}
The argument $iu$ is meant as the continuous deformation of $u\to iu$
at fixed large $|u|$. Because of the $\ln(u)$ in (\ref{maganiso}),
there are non-trivial monodromies in $M_S(u)$ as a function of $u$, so
for example $M_S(e^{2\pi i} u)$ is not necessarily equal to
$M_S(u)$. The quantum-group truncation (\ref{truncT}) also simplifies
in the zero temperature limit to
\begin{equation}
M_{Q/2}(u)=Q/2.
\label{trunc}
\end{equation}
These relations are the analog of the Seiberg-Witten monodromy
relations at infinity.

\section{The Result}

In this section we will show that the magnetization of the spin-$S$ Kondo
problem is given by
\begin{equation}
M_S(u)= \frac{iu}{4\pi}\int_{\cal C_S} \frac{dx}{xy}
\label{bigone}
\end{equation}
where
\begin{equation}
y^2 = (-1)^{2S}(x-x^g) +  u^2
\label{ykondo}
\end{equation}
where the contour surrounds the ``first'' $2S$ branch points in a
fashion to be described.

The proof of this formula follows indirectly from the integral formula
(\ref{maganiso}) and the fusion relations (\ref{fusion}). Using
(\ref{maganiso}), the magnetization can be expanded in a power series
around $u\to\infty$ (close to the weakly-coupled unstable fixed point)
and around $u=0$ (the strongly-coupled stable fixed point), each
series with a finite radius of convergence (except for $g=1$, which
will be discussed in section 5). The large-$u$ series is
(correcting a few typos in \cite{TW})
\begin{equation}
M_{1/2}(u) = \frac{1}{2\sqrt{\pi}} \sum_{n=0}^{\infty}
\frac{(-1)^n}{n!}
\frac{\Gamma(\frac{1}{2} + n(1-g))}{\Gamma(1 -ng)}
u^{-2n(1-g)}
\label{UVexp}
\end{equation}
while the small-$u$ expression is
\begin{equation}
M_{1/2}(u) = \frac{1}{\sqrt{\pi}} \sum_{n=0}^{\infty}
\frac{(-1)^n}{n!(2n+1)}
\frac{\Gamma(1 + (n+1/2)\frac{1}{(1-g)})}{\Gamma(1+
(n+1/2)\frac{g}{(1-g)})}
u^{2n+1}.
\label{IRexp}
\end{equation}
The non-trivial monodromies in the magnetization as the physical
parameter $u$ is varied in the complex plane are apparent in these
expansions: taking $u\to e^{2\pi i} u$ at fixed $|u|$ in (\ref{UVexp})
does not bring the magnetization back to its original value.

The integral expression (\ref{bigone}) follows from either expansion
by using the integral representation
\begin{equation}
\frac{\Gamma(a)}{\Gamma(a+b)} = \Gamma(1-b)\frac{i}{2\pi}
\int_{{\cal C}_{01}} dt\ t^{a-1} (t-1)^{b-1}.
\label{betaid}
\end{equation}
The integration contour ${\cal C}_{01}$ is not closed but rather starts at the
origin, loops around the branch point at $t=1$, and returns to the
origin. This formula is valid when Re$(a) > 0$ and $b$ is not an
integer. Using this in (\ref{UVexp}) yields
$$
M_{1/2}(u) = \frac{i}{2\pi^{3/2}} \sum_{n=0}^{\infty}
\frac{(-1)^n}{n!}\Gamma(1/2 +n)
\int_{{\cal C}_{01}} dt\ t^{n(1-g) -1/2} (t-1)^{-n-1/2}
u^{-2n(1-g)}.
$$
Since the sum is absolutely convergent, we can do the sum before the integral.
The sum is of the form
$$\sum_{n=0}^{\infty} \frac{(-1)^n}{n!} \Gamma(1/2 +n)z^n =
\frac{\sqrt{\pi}}{\sqrt{1+z}}.$$
This yields
$$M_{1/2}(u)=\frac{i}{4\pi}\int_{{\cal C}_{0,branch}} \frac{dt}{t^{1/2}}
\frac{1}{(t-1+t^{1-g}u^{-2(1-g)})^{1/2}}$$
where the contour ${\cal C}_{0,branch}$ starts at the origin, loops
around the square-root branch point on the positive real $t$-axis and
returns to the origin. To obtain the result (\ref{bigone}) for
$S=1/2$, one defines $x=u^2/t$ and changes variables. The
magnetization for a spin-$1/2$ impurity is therefore
\begin{equation}
M_{1/2}(u)=\frac{i}{4\pi}\int_{{\cal C}_{1/2}} \frac{dx}{x}
\frac{u}{(x^g-x+u^2)^{1/2}}.
\label{spinhalf}
\end{equation}
The contour ${\cal C}_{1/2}$ is displayed in figure 1. It starts at
$\infty$, loops around the square-root branch point along the positive
real axis and returns to $\infty$. Such a branch point must exist for
real $u$ because $y^2=x^g-x+u^2$ is negative for $x$ large but
positive for $x=0$. For complex $u$ we define ${\cal C}_{1/2}$ as the
continuous deformation of this curve away from $u$ real.
\begin{figure}
\begin{center}
{\includegraphics[scale=0.8]{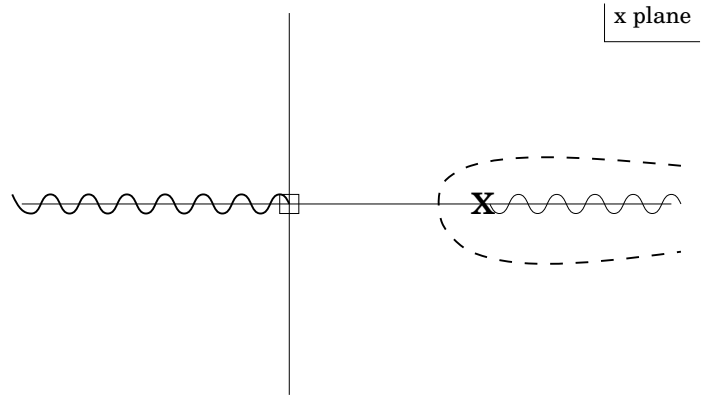}}
\caption{The dotted line is the integration contour for $M_{1/2}$ in
the $x$-plane. The box is the simple pole, the X the square-root
branch point on the first sheet of $x$, and the wavy lines are the
square-root branch cut and the $x^g$ cut from the origin.}
\end{center}
\end{figure}

This integral expression for the magnetization is much simpler than
the original result (\ref{maganiso}). This expression is analytic in
$u$ except at a few singularities. It is thus an exact,
non-perturbative expression for the magnetization.  A singularity
can occur at values $u=u_j$ where two roots coincide. This
happens when both $x^g-x+u_j^2=0$ and $gx^{g-1}-1=0$, so
\begin{equation}
u_j^2 = g^{g/(1-g)}(g-1)e^{i2\pi j/(g-1)}
\label{sing}
\end{equation}
where $j$ is some integer. All possible singularities are on the circle
$|u|=|u_j|$; for $g$ irrational they are dense along this circle.
There are no values of $u$ where three or more roots coincide.  The
integral (\ref{spinhalf}) does not diverge at all of the values
$u=u_j$, but only at values where the contour ${\cal C}_{1/2}$ goes in
between the two coinciding roots. It is easy to check that this will
never happen for real values of $u$ (where the couplings are
``physical'').  These singularities are the Lee-Yang zeroes of the
partition function, but there is no obvious physical interpretation of
them here like in gauge theory. In gauge theory, all complex values of
$u$ are physical, and the singularities correspond to couplings where
some particle becomes massless \cite{SW}.

Since it is analytic for $u$ real and positive, the integral
expression (\ref{spinhalf}) must reproduce not only the weak-coupling
series expansion (\ref{UVexp}) from which it was constructed, but also
the strong-coupling series expansion (\ref{IRexp}).  For $|u|<|u_j|$,
the contour ${\cal C}_{1/2}$ can be deformed so that $|x-x^g|>|u|^2$
for all values of $x$. Expanding the square root in powers of $u^2$
yields
$$
M_{1/2}(u) = \frac{1}{4\pi^{3/2}} \sum_{n=0}^{\infty}
\frac{(-1)^n}{n!}\Gamma(1/2 +n)
\int_{{\cal C}_{1/2}} \frac{dx}{x} (x^g-x)^{-(n+1/2)}
u^{2n+1}.
$$
Defining a new variable $t=x^{g-1}$ and using the identity
(\ref{betaid}) yields the expansion (\ref{IRexp}).  The value
$u=|u_j|$ is therefore the value at which the perturbation expansions
diverge; i.e. (\ref{UVexp}) converges for $|u|>|u_j|$ whereas
(\ref{IRexp}) converges for $|u|<|u_j|$. The small-$u$ expansion is
like the strong-coupling instanton expansion in gauge theory.

There are two kinds of branch points in the integrand; a square-root
branch cut at every root of $x-x^g-u^2$, and another at the origin due
to the $x^g$ term. If desired, the former can be removed by defining a
new variable $t=\ln x$, leaving only a simple pole at the origin. The
square-root branch points cannot be removed by a reparamaterization of
$x$. They are very interesting, because the higher-spin magnetizations
can be expressed by contours around these branch points.

The higher-spin magnetizations are found from the spin-$1/2$
magnetization by using the fusion formula (\ref{fusion}). This
requires analytically continuing the spin-$1/2$ magnetization into the
complex-$u$ plane.  How to do this is most easily seen by first
studying rational values of $g$, where $g=P/Q$ for mutually prime
integers $P$ and $Q$. The branch cut from the origin can be removed by
defining a new variable $r=x^Q$, yielding
\begin{equation}
M_{1/2}(u)= \frac{iQ}{4\pi}\int_{{\cal C}_{1/2}} \frac{dr}{r}
\frac{u}{(r^P - r^Q + u^2)^{1/2}}.
\label{rat}
\end{equation}
When $|u|$ is large, the roots $r_k$ of the polynomial $r^P-r^Q+u^2$
are approximately
$$r_k\approx e^{i2\pi k/Q}u^{2/Q}$$
for $k=0,\dots,Q-1$. The roots $r=\tilde{r}_{k}$ of the polynomial
$r^p-r^Q-u^2$ at large $|u|$ are
\begin{equation}
\tilde{r}_{k}\approx e^{i\pi (2k+1)/Q}u^{2/Q}
\label{rapprox}
\end{equation}
for $k=0,\dots,Q-1$. Therefore under the continuation $u\to i u$ at
fixed large $|u|$, the root $r_k$ moves to $\tilde r_k$, while under
$u\to -i u$, $r_k$ moves to $\tilde r_{k-1}$. In particular, the root
$r_0=u^{2/Q}$ on the real axis for real $u$ moves to
$|u|^{2/Q}e^{\pm i\pi/Q}$ under $u\to \pm i u$. The contour ${\cal
C}_{1/2}$ in (\ref{rat}) rotates accordingly.

Now it is clear how to utilize the fusion relation (\ref{fusion}) to
obtain higher spins. The spin-$1$ magnetization is
\begin{eqnarray}
M_1(u)&=& M_{1/2}(iu) + M_{1/2}(-iu) \nonumber\\
&=&\frac{iQ}{4\pi}\int_{{\cal C}_{\infty\tilde{0}}}
\frac{iu}{(r^P-r^Q-u^2)^{1/2}} +
\frac{iQ}{4\pi}\int_{{\cal C}_{\infty\widetilde{Q-1}}}
\frac{-iu}{(r^P-r^Q-u^2)^{1/2}} \nonumber
\end{eqnarray}
where the contour ${\cal C}_{\infty\tilde{k}}$ starts at infinity and
loops around the branch point at $\tilde{r}_k$. Up to a minus sign,
the integrand is the same for both pieces, so the two contours can be
subtracted from each other. This yields a closed contour encircling
both branch points, as displayed in figure 2.
\begin{figure}
\begin{center}
{\includegraphics[scale=0.8]{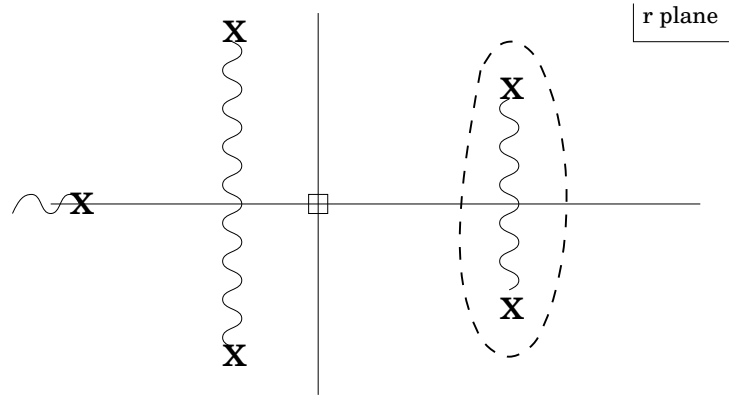}}
\caption{The dotted line is the integration contour for $M_{1}$ in the
$r$ plane for $g=4/5$. The box is the simple pole, an X is
a square-root branch point, and the wavy lines are the square-root
branch cuts. This schematic drawing is for large $u$; at $u=|u_j|$ the
two branch points on the right meet at the real axis, and for
$u<|u_j|$ they both lie on the real axis.}
\end{center}
\end{figure}

As $u$ is decreased,
the roots $\tilde{r}_k$ no longer obey the approximate relation
(\ref{rapprox}).  It follows from (\ref{sing}) that when $u=|u_j|$
these two branch points coincide at the value $x=g^{1/(1-g)}$. The
integral does not diverge because the contour is not trapped between
these two roots. Since $M_1$ is analytic at $u=|u_j|$ it can be
continued to $u<|u_j|$ where these two roots are no longer complex
conjugates but are both on the real axis. The contour ${\cal C}_1$
still just loops around the two. In fact, at value $u=|u_j|$ where
these roots coincide, the integral can be done by residue. This yields
the amusing relation
\begin{equation}
M_1\left(g^{1/(2-2g)}(1-g)^{1/2}\right) = \frac{1}{\sqrt{2g}}.
\label{amusing}
\end{equation}
The value $u=|u_j|$ can be thought of as the crossover point, the
limit of both the weak-coupling and strong-coupling series'
applicability. It is ironic that for spin $1$, the magnetization takes
on a simple value at this point.

In terms of the original integration variable $x$,
\begin{equation}
M_1(u)=\frac{1}{4\pi}\int_{{\cal C}_1} \frac{u}{(x^g-x-u^2)^{1/2}}
\label{spinone}
\end{equation}
where the contour ${\cal C}_1$ surrounds the ``first'' two branch
points.  More precisely, the first two branch points are those which
approach $x\approx \pm iu$ for $u$ large, coincide at $u=u_j$, and
approach $0$ and $1$ at $u$ small.

The magnetizations for higher spins are built up from $M_{1/2}$ in a
similar manner. The fusion relation can be rewritten as
\begin{equation}
M_S(u)=\sum_{j=-S+1/2}^{S-1/2} M_{1/2}(e^{i\pi j}u)
\label{fusiontwo}
\end{equation}
where $M_{1/2}(e^{i\alpha}u)$ means $M_{1/2}(u)$ continued from $u\to
e^{i\alpha} u$ at fixed large $|u|$.  All half-integer-spin
magnetizations therefore utilize the original curve $y^2= x^g-x+u^2$,
while the integer-spin ones use $y^2=x^g-x-u^2$. The integration
contours encircle the first $2S$ branch points. For example,
(\ref{fusiontwo}) gives
$$M_{3/2}(u)= M_{1/2}(u) + M_{1/2}(e^{i\pi}u) + M_{1/2}(e^{-i\pi}u)$$
At large $u$, continuing $u\to e^{i\pi}u$ means that the contour
starts at infinity and goes around $r_1$. Because of the relative $-$
sign, the contour for $M_{1/2}(u)+M_{1/2}(e^{i\pi} u)$ starts at
infinity and surrounds both $r_0$ and $r_1$. Adding $M_{1/2}(e^{-i\pi}
u)$ means that the contour surrounds $r_{Q-1}$ as well. This is
illustrated in figure 3. In terms of the original variable $x$, the
contour $C_{3/2}$ surrounds the ``first'' three branch points.
In general, the contour ${\cal C}_S$ is the contour
which surrounds the first $2S$ branch points but does not go around
the origin. This definition completes the proof of (\ref{bigone}).
\begin{figure}
\begin{center}
{\includegraphics[scale=0.8]{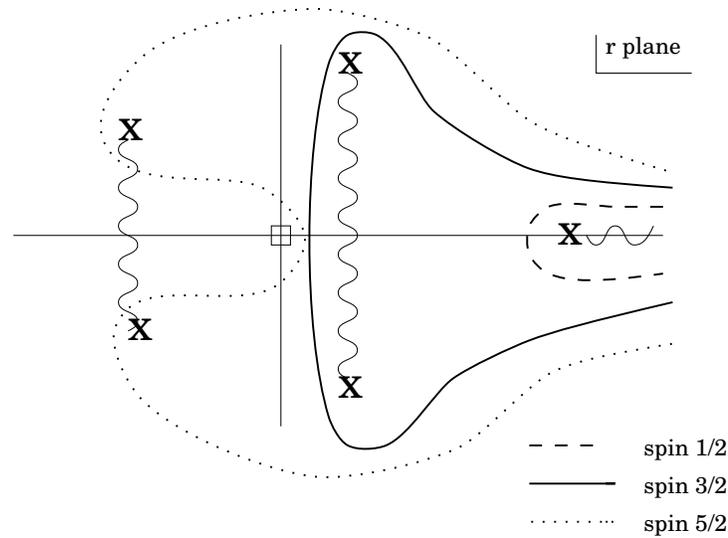}}
\caption{The contours for $M_{1/2}$, $M_{3/2}$ and $M_{5/2}$ in the
$r$-plane for $g=4/5$. The box is the simple pole, the X the
square-root branch point, and the wavy lines are the square-root
branch cuts.}
\end{center}
\end{figure}

However, one remaining question must be answered. For $g$ rational,
there are only $Q$ branch points. What are the ``first'' $2S$ branch
points when $2S>Q$? Defining $x=r^Q$ and changing variables as in
(\ref{rat}) yields
\begin{equation}
M_{S}(u)= \frac{iQ}{4\pi}\int_{{\cal C}_S} \frac{dr}{r}
\frac{u}{((-1)^{2S}(r^Q - r^P) + u^2)^{1/2}}.
\label{ratS}
\end{equation}
When $2S=Q$, the contour surrounds all the square-root branch points
but does not surround the simple pole at the origin as shown in figure
3. The contour can therefore be deformed to surround {\it only} the
origin and none of the branch points. The integral is then easily done
by residue, yielding $M_{Q/2}=Q/2$. This is precisely the
quantum-group truncation from (\ref{trunc})! The fusion relations
still hold for $S>Q/2$, but all the magnetizations can be expressed in
terms of lower-spin ones.  Likewise, the contours ${\cal C}_S$ for
$S>Q/2$ are expressed in terms of the lower-spin contours plus the
contour surrounding the origin. The latter contributes only multiples
of $Q/2$ to the magnetization.

The equation $y^2=(-1)^{2S}(r^Q-r^P) + u^2$ defines a hyperelliptic
curve, which is a Riemann surface of genus $(Q-1)/2$ for $Q$ odd and
$(Q-2)/2$ for $Q$ even. The contours ${\cal C}_S$ can be decomposed
into combinations of contours on the surface. Thus the quantum-group
truncation allows the associated surface to be of finite genus. It
seems very likely that there is a deep reason for this. However, there
is one obvious puzzling question. When $g=2/3$ (which, perhaps not
surprisingly, is where the anisotropic Kondo model is supersymmetric
\cite{nick}), the surface is a torus.  In this context, what is the
meaning of the modular parameter $\tau$ of the torus? $M_{3/2}$ is
trivial here, so for a fixed surface only $M_{1/2}$ is non-trivial
($M_1$ is related to a different surface). In other words, is there
any physical interpretation of the contour integral around the other
cycle of the torus (the integral surrounding the branch points $r_0$
and $r_1$)?

\section{$g\to 1/g$ Duality  and The Boundary Sine-Gordon Model}

In this section I discuss the $g\to 1/g$ duality present in the
anisotropic Kondo model, and the self-duality of the boundary
sine-Gordon model.

The boundary sine-Gordon model is another integrable $1+1$-dimensional
field theory defined on the half-line. It is deeply related to the
anisotropic Kondo model \cite{FLeS,BLZii}, but here there is no extra
boundary degree of freedom like the impurity spin. The bulk Lagrangian
is still (\ref{lzero}), but the boundary Lagrangian is
\begin{equation}
{L_{BSG}=2 v\cos\phi(0).}
\label{lbsg}
\end{equation}
In this model, the parameter $H/g$ is replaced by the voltage $V$,
while $v$ is related to a boundary scale $T_B$ like $\lambda$ is
related to $T_K$. The analog of the magnetization is the current $I$
arising from the applied voltage. The power-series
expansions for the normalized current ${\cal I}=I/gV$
as a function of $u\propto Vv^{-1/(1-g)}$
are \cite{FLSbig}
\begin{equation}
{\cal I}(u,g)= 1 - \sum_{n=1}^\infty a_n(g) u^{2n(g-1)}
\label{bUV}
\end{equation}
and
\begin{equation}
{\cal I}(u,g)= \sum_{n=1}^\infty a_n(1/g) u^{2n(1/g-1)}
\label{bIR}
\end{equation}
where
\begin{equation}
a_n(g) = \frac{(-1)^n}{n!} \frac{\sqrt{\pi}}{2}
\frac{\Gamma(ng+1)}{\Gamma(3/2 +n(g-1))}.
\label{asubn}
\end{equation}
This model is self-dual because the weak-coupling expansion is
identical to the strong-coupling expansion by the exchange
$g\to 1/g$. This leads to a simple statement about the
behavior of the current as an analytic function of $g$ \cite{FLSbig}:
\begin{equation}
{\cal I}(u,g) = 1 - {\cal I}(u,1/g)
\label{selfdual}
\end{equation}

A simple integral expression for the current can be derived from
either expansion (\ref{bUV}) or (\ref{bIR}). The proof is similar to
that given in section 3 for the Kondo model. The result is
\begin{equation}
{\cal I}(u,g)=\frac{i}{4u}
\int_{{\cal C}_B} dx \frac{1}{\sqrt{x + x^g - u^2}}
\label{Ilutt}
\end{equation}
where the curve ${\cal C}_B$ starts at the {\it origin}, loops around
the branch point on the positive real $x$-axis, and returns to the
origin. Proving the self-duality relation (\ref{selfdual}) using
(\ref{Ilutt}) is easy because this integral expression is valid for
all values of $g$: start with ${\cal I}(u,1/g)$, change variables
$x\to x^g$, and integrate by parts.  The self-duality of the boundary
sine-Gordon model and some generalizations is discussed in detail in
\cite{FS}. In the context of the fractional quantum Hall effect, this
duality exchanges electrons with Laughlin quasiparticles.

The $g\to 1/g$ duality in the Kondo model is not so
obvious. Nevertheless, it was noted some time ago \cite{FW} that the
isotropic Kondo model has a duality relating the large-$u$ expansion
for spin $S$ to a piece of small-$u$ expansion for spin $S+1/2$. The
shift in $S$ is not shocking because at the strongly-coupled ($u=0$)
fixed point, one of the electrons binds to the impurity, effectively
reducing the spin from $S+1/2$ to $S$. This duality is not a
self-duality, because it maps one spin onto another. To see how this
duality arises using the integral (\ref{bigone}), first note that if
one changes variables $x\to x^{1/g}$, one has
\begin{equation}
M_S(u)=\frac{i}{4\pi g}\int \frac{dx}{x}
\frac{u}{((-1)^{2S+1}(x^{1/g}-x)+u^2)^{1/2}}.
\label{kdual}
\end{equation}
where the contour must be defined.  Up to the extra factor of $g$ in
front, the integrand is exactly that for the magnetization with $g$
replaced with $1/g$, and integer spins exchanged with
half-integer. However, a simple formula like (\ref{selfdual}) does not
instantly follow because the contour in (\ref{kdual}) must be
carefully defined.

The subtlety involving the dual contours can be seen in the small-$u$
expansion of $M_1(u)$. As discussed in the previous section, for
$u<|u_j|$ the contour ${\cal C}_1$ loops around the two roots on the
positive real axis. For $u$ small, these roots of $x-x^g+u^2$ are at
$x\approx u^{2/g}$ and at $x\approx 1$. The contour ${\cal C}_S$ can
be decomposed into two contours: one starting at infinity and looping
around the root near $x=1$, and the other starting at infinity and
looping around the root near $x=u^{2/g}$. The integral can therefore
be split into two pieces. The first piece is like $M_{1/2}(u)$, but
the integrand is slightly different because of the $(-1)^{2S}$. For
small $u$ this piece can still be expanded in powers of
$u^{2n+1}$ (the resulting coefficients are in fact $i(-1)^n$ times
those in (\ref{IRexp}); this piece is purely imaginary).  The duality
is seen in the second piece, which is denoted $\widetilde
M_1(u)$. Defining a new variable $t=x^g/u^2$ gives
\begin{equation}
\widetilde M_1(u)=\frac{i}{4\pi g}\int_{{\cal C}_{\infty 1}}
\frac{dt}{t} \frac{1}{(u^{-2+2/g}t^{1/g}-t+1)^{1/2}}.
\label{MIR}
\end{equation}
The contour ${\cal C}_{\infty 1}$ loops around the branch point near
$t=1$. For small $u$ and fixed $t$, the square root can be expanded in
powers of $u^{(-2+2/g)}$:
$$
\widetilde M_1(u) =
\frac{i}{4\pi^{3/2} g}\int_{{\cal C}_{\infty 1}}\frac{dt}{t}
\sum_{n=0}^\infty \frac{(-1)^n}{n!}\Gamma(n+1/2)
t^{n/g}(1-t)^{-n-1/2}u^{2n(1/g-1)}.
$$
The resulting integral in $t$ converges if $n<g/(2(1-g))$ (or
equivalently $u^{2n(1/g-1)} <u$), but the terms for higher $n$ can
formally be defined by continuing in $g$ from values close to $1$. The
integrals for larger $n$ diverge because integer powers $u^n$
(possibly multiplying $\ln(u)$) also contribute to the small
$u$-expansion of $\widetilde M_1(u)$. (This can be seen explicitly by
breaking ${\cal C}_{\infty 1}$ into two pieces so that one can
construct convergent expansions.) The integer powers must cancel the
imaginary pieces from the first contour to ensure that $M_1(u)$ is
real for real positive $u$.  By using the identity (\ref{betaid}), the
magnetization is
\begin{equation}
M_{1}(u) = \frac{1}{2\sqrt{\pi}g} \sum_{n=0}^{\infty}
\frac{(-1)^n}{n!}
\frac{\Gamma(\frac{1}{2} + n(1-1/g))}{\Gamma(1 -n/g)}
u^{-2n(1-1/g)} + {\cal O}(x^n).
\label{kondual}
\end{equation}
The ${\cal O}(x^n)$ terms include the contributions from the first
contour as well, and also include terms like $x\ln x$.  The duality in
the Kondo model is now apparent: the contributions to $M_1(u)$ at
small $u$ displayed in (\ref{kondual}) are precisely those in the
large-$u$ expansion of $M_{1/2}(u)$ in (\ref{UVexp}), with the
replacement $g\to 1/g$, up to the extra factor of $g$. Thus for
example, we recover the fact that $M_1(0)=1/(2g)$, as mentioned
before.

The proof of the duality for general $S$ is straightforward.  For
example, the contour ${\cal C}_{3/2}$ can be split into two contours,
one around the branch point on the positive real axis, and the other
surrounding just the first two complex-conjugate branch points
(e.g.\ $r_1$ and $r_{Q-1}$ for rational $g=P/Q$). The first contour
gives merely $M_{1/2}(u)$. The second contour is best understood by
using the variable $t=x^g/u^2$ just like in (\ref{MIR}). In the
small-$u$ limit, the second contour surrounds the square-root branch
point $t=-1$ just above the branch cut from the origin and and the
square-root branch point $t=-1$ just below the branch cut from the
origin. This contour is the same as the contour for $M_1(u)$ in the
large-$u$ limit, and moreover, the integrand is the same as
$M_1(u)$ for large $u$, up to the replacement $g\to 1/g$ and with an
overall $1/g$. In other words,
\begin{eqnarray}
M_{3/2}(u)&=&\frac{i}{4\pi}\int_{{\cal C}_{3/2}} \frac{dx}{x}
\frac{u}{(x^g-x+u^2)^{1/2}}
\nonumber\\
&=&M_{1/2}(u)+\frac{i}{4\pi g}\int_{{\cal C}_{1}} \frac{dt}{t}
\frac{1}{(t-u^{-2+2/g}t^{1/g}+1)^{1/2}}
\nonumber\\
&=&M_{1/2}(u)+ \frac{1}{g} M_1(u)\bigg|_{g\to 1/g}
\end{eqnarray}
By this argument, for integer $S$ one has
\begin{equation}
M_{S+1/2}(u)=M_{1/2}(u) + \frac{1}{g} M_S(u)\bigg|_{g\to 1/g}
\end{equation}
where the $g\to 1/g$ in the last line is precisely defined at least at
small $u$.  For half-integer $S$, such a simple relation does not seem
to be true, but by these arguments
\begin{equation}
M_{S+1/2}(u)=\frac{1}{g} M_S(u)\big|_{g\to 1/g}+ {\cal O}(u^n)
\end{equation}
is true for all $S$. This is the Kondo duality discussed in
\cite{FW}. Given the integral relation (\ref{kdual}) it seems likely
that this is not the only form of $g\to 1/g$ duality in the Kondo
model.

\section{The SU(2) Point}

A simple integral expression for the zero-temperature magnetization of
the anisotropic Kondo model was derived in the section 3. The results
most resemble the Seiberg-Witten results for gauge theory when $g$ is
rational and less than $1$, and the integral is on a
finite-genus hyperelliptic curve. However, the model which is most
like QCD is the $SU(2)$-symmetric case $g=1$. As explained in section
2, the Kondo model is asymptotically free, develops a mass scale even
though all parameters are naively dimensionless, and exhibits a
crossover into a strong-coupling phase. In this section I discuss the
isotropic Kondo model in more detail, find explicit expressions for
its perturbative coefficients, and highlight its $g\to 1/g$ duality.

Taking the limit $g\to 1$ of (\ref{bigone}) yields the appropriate
integral expression. The definition (\ref{udef}) shows that for fixed
$H/T_K$, $u^2\to (1-g) H^2/(2\pi T_K^2)$ as $g\to 1$, so
$$\lim_{g\to 1}\ x^g-x+u^2 = (g-1)\left(x\ln x-H^2/(2\pi T_K^2)\right).
$$
Denoting ${\cal M}_S(H/T_K)=\lim_{g\to 1} M_S(u)$, its integral form is
\begin{equation}
{\cal M}_S(H/T_K)=\frac{i}{4\pi}\int_{{\cal C}_S} \frac{dx}{x}
\frac{H/T_K}{((-1)^{2S}2\pi x\ln x + (H/T_K)^2)^{1/2}}
\label{iso}
\end{equation}
The contour ${\cal C}_S$ is defined as in the previous section; there
are of course an infinite number of solutions to the transcendental
equation $2\pi x\ln x =\pm (H/T_K)^2$, so as expected for $SU(2)$
symmetry, all the ${\cal M}_S$ are independent.

Weak- and strong-coupling perturbative expansions can be derived from
(\ref{iso}), but there are some subtleties. For spin $1/2$, the small
$H/T_K$ expansion follows as before, by expanding the square root in
powers of $(H/T_K)^2$. One indeed obtains the $g\to 1$ limit of
(\ref{IRexp}). However, the form of the large-$H/T_K$ expansion (which
corresponds to the weak-coupling expansion in terms of the original
coupling $\lambda$) is not so obvious: the coefficients in
(\ref{UVexp}) diverge as $g\to 1$, while the powers all collapse to
zero. As seen from the definition (\ref{tklam}) of $T_K$ at the
$SU(2)$ point, the perturbative expansion in $\lambda$ should include
terms with $\log T_K$. This is the usual behavior of an
asymptotically-free theory. This suggests defining a renormalized
coupling $z(H,\lambda,D)$ ($D$ is the cutoff) which is
renormalization-group invariant and which obeys $z\approx \lambda$ for
$\lambda\ll 1$.  The explicit perturbative results (summarized in
\cite{TW}) suggest
\begin{equation}
\ln(H/T_K)=\frac{1}{z}-\frac{1}{2}\ln (z/4\pi).
\label{zdef}
\end{equation}
This differs from the definition in \cite{TW} by a $z$-independent
numerical factor on the right; the reason will be apparent shortly.

Weak coupling ($H/T_K$ large) is equivalent to $z$ small.
The magnetization $M_{1/2}(u)$ can be expanded in a power series in $z$:
\begin{eqnarray}
{\cal M}_{1/2}(z)&=& \frac{i}{4\pi}\int_{{\cal C}_{1/2}} \frac{dx}{x}
\frac{\sqrt{2}}{(- x(\ln x + \ln(H^2/4\pi T_K^2)) +2)^{1/2}}
\nonumber\\
&=& \frac{i}{4\pi}\int_{{\cal C}_{1/2}} \frac{dx}{x}
\frac{\sqrt{2}}{(- x(\ln x + 2/z - \ln z) +2)^{1/2}}
\nonumber\\
&=& \frac{i}{4\pi}\int_{{\cal C}_{1/2}} \frac{dx}{x}
\frac{1}{(- (z/2)x\ln x -x +1)^{1/2}}
\label{magz}
\end{eqnarray}
where the first line required a rescaling $x\to xH^2/(4\pi T_K^2)$ and the
third line required $x\to zx$. Now the square root can be expanded in
powers of $z$:
\begin{equation}
{\cal M}_{1/2}(z)=\sum_{n=0}^\infty{\cal A}_n z^n
\label{expz}
\end{equation}
where
$${\cal A}_n=\frac{(-1)^n}{2^{n+1}\pi^{3/2} n!}\Gamma(n+1/2)
\int_1^\infty dx\ x^{n-1} (\ln x)^n (x-1)^{-n-1/2}.
$$
These integrals converge for all $n$, but with a different choice of
the constant in (\ref{zdef}) they would not.  This expansion in $z$ is
asymptotic: for large enough $x$ the term $zx\ln x$ eventually
dominates the other terms in (\ref{magz}) no matter how small $z$ is,
so the expansion in powers of $z$ has zero radius of convergence.  Of
course this is expected for a theory with dimensional transmutation;
what is remarkable here is that even the non-perturbative corrections
are included in the Bethe ansatz solution of the Kondo problem and in
the simple integral expression (\ref{magz}).  The leading term is
${\cal A}_0=1/2$ as required, and ${\cal A}_1=-1/4$. The integral
cannot be done in closed form for higher $n$, although it can be
expressed in term of a hypergeometric function at fixed
argument. Numerically, ${\cal A}_2=0.298286794$, ${\cal
A}_3=-0.648160191, \dots$, and as $n\to\infty$, ${\cal A}_n \propto
n!$.

The last remaining issue is how the duality manifests itself at the
$SU(2)$ point $g=1$. To find a perturbative expansion for ${\cal M}_1$
for small $H/T_K$, it is convenient to define the parameter $\tilde z$
analogously to $z$:
\begin{equation}
\ln(H/T_K)=-\frac{1}{\tilde z}-\frac{1}{2}\ln (\tilde z/4\pi).
\label{ztilde}
\end{equation}
The changed sign in (\ref{ztilde}) means that $\tilde z\to 0$ as
$H/T_K\to 0$ (notice that $\tilde z$ is imaginary for large $H/T_K$).
The spin-$1$ magnetization is then
\begin{equation}
{\cal M}_1(\tilde z)=\frac{i}{4\pi}\int_{{\cal C}_{1}} \frac{dx}{x}
\frac{1}{((\tilde z/2)x\ln x -x +1)^{1/2}}
\end{equation}
For $z$ small the contour surrounds the branch points at $x\approx
2/z$ and $x\approx 1$. As for $g<1$, the contour can be split into two
pieces, the first extending from infinity around the branch point near
$2/z$, and the second extending from infinity around the branch point
near $1$. The first can be expanded in a power series in $(H/T_K)^n$,
as can easily be seen by returning to the original integral
expression. Denoting the second piece $\widetilde{\cal M}_1(\tilde
z)$, it follows immediately that
$$\widetilde{\cal M}_1(\tilde z)={\cal M}_{1/2}(-z)$$
The expansion of $\widetilde{\cal M_1}(\tilde z)$ in powers of $\tilde
z$ is asymptotic. Putting the two contours back together gives
\begin{equation}
{\cal M}_1(\tilde z)={\cal M}_{1/2}(-z) + {\cal O}\left((H/T_K)\right).
\end{equation}
In this sense the parameter $\tilde z$ is dual to $z$, just like the
earlier Kondo duality required sending $g\to 1/g$. It is possible that
there is a better definition of $z$ and $\tilde z$ which makes the
duality relation more transparent.

\section{Conclusions and Questions}

I have shown that the zero-temperature magnetization in the spin-$S$
Kondo model can be written as integrals around cycles of a curve
$y^2=(-1)^{2S}(x-x^g)+u^2$. At the $SU(2)$ point, this curve reduces
to $y^2=x\ln x -H^2/2\pi T_K^2$. At rational $g<1$, this curve is of
finite genus, a fact connected to the quantum-group symmetry of the
model. The curve also gives a simple way to see the duality in the
Kondo problem between the weak-coupling spin-$S$ problem and the
strong-coupling spin-$(S+1/2)$ problem.

The Kondo model is integrable, which implies the existence of an
infinite-dimensional symmetry. The results in this paper clearly
utilize the integrability of the model, so even though the Kondo model
is not supersymmetric in general it does possess a great deal of
symmetry. In fact, the infinite-dimensional quantum-group symmetry
underlying some integrable models can be shown to be an extension and
deformation of two-dimensional $N=2$ supersymmetry (which exists at
the point $g=2/3$ here) \cite{BL}. The question is then if such a
deformation is possible in four dimensions: for example, can the
$SU(2)$ gauge symmetry of \cite{SW} be deformed into some sort of
quantum-group-like symmetry, which may break the supersymmetry but
still enable the Seiberg-Witten computations? In fact, by deforming
away from a supersymmetric point, it is known that duality holds in
non-supersymmetric string theories \cite{BD}.

A potentially related issue is the fact that the results of
\cite{SW,SWreview} are known to be intimately connected to classical
integrable systems \cite{integ}. The models discussed in this paper
are quantum integrable systems. There is no obvious relation between
the two, but it is hard to imagine that there is none.

Since the Seiberg-Witten results follow from a finite-dimensional
symmetry algebra, this raises a question for the two-dimensional
models: is the full power of integrability necessary for these
computations to be valid, or is some smaller symmetry algebra
sufficient? For example, in two dimensions at least one way of
exploiting a finite-dimensional symmetry algebra is known, the
topological-antitopological fusion of \cite{CV}. The results here hint
that there may be a completely new way of approaching
$1+1$-dimensional field theories, integrable or not.  One way of
approaching the problem is to find a physical interpretation of the
monodromies around the singularities; the fusion relation
(\ref{fusion}) says something about the monodromies at infinity, but
not much about those at $|u|=|u_j|$. Of course, since the exact
solution is known these monodromies can be found, but it would be much
more desirable to know them {\it a priori} instead of {\it a
posteriori}.

Another important direction to explore is to understand if these
results can be extended to finite temperature, where the fusion
relations (\ref{fusion}) are nonlinear. This may seem to destroy the
whole picture, since the addition of contours is a linear
relation. However, the current in the boundary sine-Gordon model is
also related to the magnetization in Kondo by a non-linear fusion
relation \cite{FLSbig,BLZii}, and still obeys the integral expression
discussed in section 4 and in \cite{FS}.

The Kondo model is one of the grand old problems of quantum
statistical mechanics. It was recognized long ago to be a useful toy
model for QCD \cite{Wilson}.  It is remarkable that more than two
decades later, it still holds a few surprises.
\bigskip\bigskip 

I would like to thank K. Intriligator and P. Arnold
for many useful conversations on duality, B. McCoy for interesting comments
on the paper, and H. Saleur for related collaboration.
\bigskip\bigskip\bigskip
\renewcommand{\baselinestretch}{1}

\end{document}